\documentclass[10pt]{article}
\usepackage{graphicx,amssymb,amsmath}
\newcommand{\be}{\begin{equation}}
\newcommand{\ee}{\end{equation}}
\newcommand{\beq}{\begin{eqnarray}}
\newcommand{\eeq}{\end{eqnarray}}
\newtheorem{thm}{Theorem}[section]

\def\d{\mathrm{d}}

\def\ben{\begin{equation}}
\def\een{\end{equation}}
\def\half{{1 \over 2}}
\def\bea{\begin{eqnarray}}
\def\eea{\end{eqnarray}}
\def\ft#1#2{{\textstyle{\frac{\scriptstyle #1}{\scriptstyle #2}}}}
\def\fft#1#2{\frac{#1}{#2}}

\begin{document}

\title{Metrics With Vanishing Quantum Corrections}

\author{ \Large A.A. Coley$^{1}$, G.W. Gibbons$^{2}$,
S. Hervik$^{1}$\footnote{Address after 15 April 2008: Dept. of Mathematics and Natural Sciences, University of Stavanger, N-4036 Stavanger, Norway } ~ and C.N. Pope$^{2,3}$
\\
\\
$^{1}$Dept of Mathematics \& Statistics,
Dalhousie University
\\Halifax N.S. B3H 3J5
Canada
\\
$^{2}$D.A.M.T.P.,
 Cambridge University,
\\ Wilberforce Road,
 Cambridge CB3 0WA,
U.K.
\\ $^{3}$ George P. \& Cynthia W. Mitchell Institute for Fundamental Physics,
\\Texas A\&M University,
College Station,
\\  TX 77843-4242,
USA
\\ \\
}

\maketitle

\begin{abstract} 
We investigate solutions of the classical Einstein or supergravity
equations that solve  {\sl any} set of quantum corrected Einstein equations
in which the Einstein tensor plus a multiple of the metric
is equated to a symmetric conserved tensor $T_{\mu \nu }(g_{\alpha
  \beta },
\partial _\tau  g_{\alpha  \beta } ,\partial  _\tau \partial _\sigma 
 g_{\alpha  \beta } \,, \dots , ) $ 
constructed from sums of terms the involving contractions of the
metric and powers of arbitrary covariant derivatives  of the  curvature
tensor.
A classical solution, such as an Einstein metric, is called
{\it universal}  if,  when evaluated on that Einstein metric, $T_{\mu \nu}$ is
a multiple of the metric. A Ricci flat classical solution
is called {\it strongly universal} if,  when evaluated on that Ricci flat
 metric, $T_{\mu \nu}$ vanishes. It is well known that pp-waves
in four spacetime dimensions are strongly universal. 
We focus attention on a natural generalisation;
Einstein metrics with holonomy ${\rm Sim} (n-2)$  in which all scalar
invariants are zero or constant. In four dimensions we demonstrate
that the generalised Ghanam-Thompson metric is weakly universal
and that the Goldberg-Kerr metric is strongly
universal; indeed, we show that universality extends to all 4-dimensional ${\rm Sim}(2)$ Einstein metrics. We also discuss 
generalizations to higher dimensions.

\end{abstract}

\vskip0.3in
\leftline{MIFP-08-04}

\vspace*{0.3in}

\thispagestyle{empty}

\pagebreak
\voffset=0pt

\setcounter{page}{1}

\tableofcontents

\addtocontents{toc}{\protect\setcounter{tocdepth}{2}}

\newpage

\section{Introduction}

It has been realised for some time that
certain  solutions of the classical  Einstein 
equations, or some variant of them such as the classical
supergravity equations, remain valid solutions
in the quantum theory, despite our ignorance of precisely
what that quantum theory might be. Indeed, some solutions
of the  classical Einstein equations have such a restricted
curvature structure that they  remain valid solutions
of {\sl almost any}  set of covariant equations involving the metric
and its derivatives. 
We shall call such metrics {\it Universal}, with a further subdivision
into {\it Strongly Universal} and {\it Weakly Universal}, which we shall
explain below.

We assume that the  field equations for an $n$-dimensional
spacetime metric take  the form
\ben
G_{\mu \nu} = - \Lambda g_{ \mu \nu} +  2 T_{\mu \nu}\,,
\label{eom} \een
where $G_{\mu \nu}= R_{\mu \nu}- \half R g_{\mu \nu} $ 
is the Einstein tensor, $\Lambda$ is a classical cosmological constant term,
and $T_{\mu \nu}$ represents possible quantum corrections to the
classical Einstein equations in the case of classical vacuum. Necessarily the 
symmetric second rank tensor    $ T_{\mu \nu} =
T_{\mu  \nu }(g_{\alpha \beta},
\partial _\tau  g_{\alpha \beta } ,\partial  _\tau \partial _\sigma 
 g_{\alpha \beta } \,, \dots , ) $,   which we assume to be  made up of 
sums of terms each  constructed from the metric,
the curvature tensor $R_{\mu \nu \sigma \tau}$ and its covariant derivatives,
is conserved; i.e., 
\ben
T^{\mu \nu} \, _{; \nu}  =0 \, .
\een

Now  suppose we have a solution $g_{\mu \nu}$  of the classical equations,
obtained by omitting the tensor $T_{\mu \nu}$ from the right hand side 
of (\ref{eom}); i.e.,  an Einstein metric   for which
\ben
R_{\mu \nu}= \frac{2}{n-2}  \Lambda g_{\mu \nu}\,.
\een 
We ask whether the  classical  metric $g_{\mu \nu}$
 (possibly rescaled by a constant factor $h$)  solves the full
quantum corrected equations (\ref{eom}).  This requires that 
\ben
T_{\mu \nu}(h g_{\rho\sigma}) = F(h) g_{\mu \nu} \label{condition1}
\een
for some function $F(h) $ and that  $h$ may be chosen to satisfy 
\ben
\Lambda (h-1)   =  2F(h) \,. \label{equation}
\een
 
A sufficient condition that (\ref{condition1}) hold
is that it  hold for {\sl any} symmetric conserved tensor constructed
from the classical metric $g_{\mu \nu}$ and its derivatives.
We call this condition {\it Weak Universality} because,
subject to there being real solutions of (\ref{equation}),
we can simply rescale the metric to get a solution of any set of corrected
field equations.      

An example  of a weakly  universal metric  is a maximally symmetric space,
such as  de Sitter spacetime, 
for which \ben
R_{\mu \nu \sigma \tau}= c
 \bigl (g_{\mu \sigma} g_{\nu \tau}- g_{\nu \sigma} g_{\mu \tau} \bigr )\,, 
\een 
where $c$ is necessarily a constant, and hence
\ben
R_{\mu \nu \sigma \tau ;{\lambda _1} ;\dots ;{\lambda _k}  }=0
\label{symmetric} 
\een 
for arbitrary integers $k$.
It follows that $c$  must satisfy 
\ben
T_{\mu \nu}= f(c) g_{\mu \nu}\,, 
\een
for some function $f(c)$, and if a value of $c$ can be found satisfying
\ben
f(c)+ { (n-1)(n-2) \over 4}c =0\,,
\een
we have a solution.
This argument can obviously  
be generalised to cover the case of any   symmetric space
$M=G/H$, where $G$ is the isometry group with stabiliser $H$, for which by 
definition 
\ben
R_{\mu \nu \sigma \tau ;{\lambda }  }=0\,,\label{symmetric2}
\een 
if, in addition, we assume that  $H$ acts irreducibly on the tangent space.
In fact, as long as $H$ acts irreducibly (\ref{symmetric2}) is not necessary;
it suffices that the space be locally homogeneous. 
Bleecker \cite{Bleecker} has called  the restricted class of 
{\sl  Riemannian}  metrics of the
sort we are considering
{\it Critical Metrics}, as long as
the field equation  may be derived from a diffeomorphism invariant   
action functional, which in our case means
\ben
 \frac 14 \int \sqrt{| g|}  \left(  R-2 \Lambda   \right)
+  I (g)  
\label{action}\een  
such that
\ben
T_{\mu \nu} = -\frac{2}{\sqrt{| g|}} 
\frac{\delta I (g)}{ \delta g^{\mu \nu} }.
\een
In what follows we shall use Bleecker's  term critical
metric  for a metric of any signature.\footnote
{Interestingly, Bleecker \cite{Bleecker2} 
also discussed the idea of critical maps between Riemannian manifolds,
which have obvious  relevance to non-linear sigma models
and string theory} 

A necessary and sufficient condition for a metric to be {\it weakly 
universal}  is that
any conserved symmetric tensor constructed from the metric
and its derivatives  should be a 
multiple of the metric.
A necessary and sufficient condition for a metric to be {\it critical} is that
any conserved symmetric tensor constructed from the metric
and its derivatives that is a  variational derivative 
of an invariant functional should be a 
multiple of the metric. Clearly a sufficient  condition for
a metric to be critical is that it be universal. 
It is not completely obvious whether or not the converse is true.

In fact, Bleecker showed that critical  compact orientable
Riemannian manifolds    
must be  homogeneous  spaces $G/H$, where $H$ acts irreducibly.
However, the metrics we usually encounter in physics are {\it Lorentzian}, 
and there are many more  possibilities.
For example, one might consider what in this paper we shall call
a vacuum Brinkmann wave \cite{Brinkmann1,Brinkmann2,Brinkmann3}.\footnote{We shall define a Brinkmann wave as a 
spacetime admitting a covariantly constant null vector. Vacuum pp-waves are examples of vacuum 
Brinkmann spacetimes.}  This is a Ricci 
flat Lorentzian metric
admitting a covariantly constant   null vector field (CCNV) $n^\nu$;
\ben
\nabla _\mu n^ \nu =0\,.
\een 
The most general metric admitting a covariantly constant null vector can be written \cite{CFHP,Ortin}
\beq
\d s^2=2\d u\left[\d v+H(u,x^k)\d u+A_i(u,x^k)\d x^i\right]+
g_{ij}(u,x^k)\d x^i\d x^j,
\label{CCNV}\eeq
and is a subclass of the Kundt metrics \cite{CSI}.
Requiring the vanishing of the Ricci tensor implies that the transverse metric $g_{ij}(u,x^k)$ is Ricci flat. 

A commonly studied subclass of this class of metrics is the class for which $g_{ij}(u,x^k)$ is independent of $u$ and flat and $A_i=0$. 
Horowitz and Steif \cite{HorowitzSteif}  
have shown that all such metrics are not only weakly universal
but they possess an even stronger property, 
which we shall call {\it Strong  Universality}.

Clearly for Brinkmann waves, since the   Ricci tensor $R_{\mu \nu}$ vanishes,
the classical value of
$\Lambda$ also vanishes. Equally clearly,  
any constant multiple $h$ of the classical metric is also a solution.
 Horowitz and Steif showed, when evaluated on a Brinkmann wave background,
all other conserved  tensors $T_{\mu \nu}(g_{\rho\sigma}) $ 
(except of course the metric itself) will vanish. Thus, in distinction
to the case of maximally symmetric spaces such as de Sitter spacetime,  
no rescaling of the metric is required when passing from the 
classical to the quantum-corrected metric.
To make the notion of strong universality precise 
we need to consider, for a general spacetime,
the vector space (over real constants)
of all symmetric conserved second rank tensors constructed from
the metric and its derivatives, modulo constant multiples of the metric
itself. If, when restricted to a classical metric $g_{\mu \nu}$ such as
a Brinkmann wave, all such tensors vanish, then we  say that the classical
metric is {\it Strongly Universal}.

A result related to that of Horowitz and Steif  has been obtained by Torre
\cite{Torre}, who showed  that any metric with the same isometry group
as plane (or Rosen) waves, but which does not necessarily 
solve the vacuum Einstein equations, is strongly universal.   
Some physical implications of these facts for quantum field theory and 
string theory may be found  in
\cite{Gibbons,Deser,HorowitzSteif}.

One of the  aims of the present paper is
to see whether these results for Brinkmann waves can be extended
to the  newly constructed class  of solutions 
of the Einstein equations \cite{GibbonsPope} 
 which admit a  null vector field $n^\nu$    
that is covariantly constant merely in direction, 
sometimes called {\it recurrent}; i.e.,
\ben
\nabla _\mu n^\nu = B_\mu n^\nu\,,
\een
for some recurrence one-form $B_\mu$. 
An equivalent way of expressing this condition is 
to say that the corresponding metrics have holonomy
contained in ${\rm Sim} (n-2)$, the maximal proper sub-group
of the Lorentz group $SO(n-1,1)$. In the Brinkman case, for which
$B_\mu =0$, the holonomy is contained in the subgroup of Euclidean 
motions ${\rm E}(n-2)$. In some cases, the holonomy reduces even further.
For example, for some simple Brinkmann metrics the holonomy group 
could be contained in the abelian subgroup 
$\mathbb{R}^{n-2}$ of ${\rm Sim}(n-2)$. 

A principal result of our  work is the discovery
of two  new classes of four-dimensional metrics,
one of  which is  strongly universal
and the other of which  is  weakly universal.

Before embarking on our discussion we wish to make a brief remark on the issue of vanishing local counterterms.
In our context the integrand of $I(g)$ in eq. (\ref{action}) would be called a  diffeomorphism 
invariant  local counterterm, and its vanishing subject to the Ricci flat
condition  $R_{\mu \nu}=0$ (i.e., \lq\lq on shell \rq \rq) is 
often taken as an indication that 
quantum corrections are finite. However, this does not necessarily mean
that quantum corrections vanish, since one can conceive of cases
for which $I(g)$ vanishes while its variational derivative
$-\half T_{\mu \nu}$ does not.

In fact, this is precisely what happens in the case of Calabi-Yau
compactifications in string theory. The specific scalar invariants
$I(g)$ that arise as higher-order string corrections to the effective action all vanish for the product of four-dimensional Minkoswski
spacetime with a manifold of $SU(3)$ holonomy, 
but their variations $\delta I \over \delta g_{\mu\nu}$ do not \cite{frepop}.
The classical  Ricci flatness condition is consequently modified by quantum
corrections, although the supersymmetry is preserved 
(see, for example, \cite{LuPopeStelle} for an extensive 
discussion).  A similar phenomenon
arises in compactifications involving manifolds
with $G_2$ and $Spin(7)$ holonomy \cite{lupostto1,lupostto2}.
 
Another example of this phenomenon is the Ricci type N, Weyl type N Brinkmann waves for which \emph{all} scalar invariants vanish \cite{CFHP}. However, for an action containing the invariant $R_{\mu\nu}R^{\mu\nu}$, its variation will contain a term $\nabla^2R_{\mu\nu}$ which may or may not vanish. For example, for the metric (\ref{CCNV}) with $u$-independent and flat $g_{ij}(u,x^k)$ with $A_i=0$, this term gives a contribution $(\nabla^2)^2H$ which will not vanish in general.  
 
    It is nevertheless of interest to ask whether all 
scalar invariants formed from  the metric and its derivatives vanish
on shell; the so-called {\it Vanishing Scalar Invariants}, or
 VSI, condition. An Einstein  metric with vanishing scalar invariants  
must, of course, be Ricci flat.  Examples of VSI spaces are all
Ricci-flat Brinkmann waves in four and five dimensions. A  weaker, 
but nevertheless still interesting and potentially important condition,
is when  all scalar invariants are constant; the so-called {\it Constant
Scalar Invariants}, or CSI, condition. For example, we can imagine 
having a symmetric tensor of the form 
\[ T_{\mu\nu}={\cal I}\, g_{\mu\nu}, \] 
where ${\cal I}$ is a curvature invariant. Requiring this tensor to be conserved, 
$T^{\mu\nu}_{\phantom{\mu\nu};\nu}=0$, immediately implies ${\cal I}_{,\mu}=0$, 
and hence that ${\cal I}$ is a constant. It is therefore natural to 
search among 
the CSI spacetimes for weakly universal spacetimes. It is not obvious 
whether or not all weakly universal spacetimes are CSI spacetimes.

To summarise: we have isolated five conditions on a metric:
Strongly Universal; Weakly Universal; Critical; 
Vanishing Scalar Invariants (VSI)
and Constant Scalar Invariants (CSI). In what follows, we shall    
study the extent to which these conditions hold for the
Einstein metrics with ${\rm Sim}(n-2)$ holonomy, and particularly the new solutions of \cite{GibbonsPope}.

\section{Previous work and the situation in four dimensions} 

As shown in \cite{GibbonsPope}, by rescaling
the null vector field $n^\mu$ of a metric with ${\rm Sim}(n-2)$ 
holonomy  we may always arrange things so that
\ben
\nabla _\mu n ^\nu = \kappa n_\mu n^\nu\,.
\een 
It follows  that the null congruence with tangent vector
$n^\mu$ is geodesic,  hypersurface orthogonal (i.e., twist-free),
expansion-free and shear-free and hence, in all spacetime dimensions, belongs to the class of
Kundt metrics \cite{CSI}. However not every Kundt metric has
holonomy  in  ${\rm Sim}(n-2)$. Metrics with ${\rm Sim}(n-2)$ 
holonomy may be cast in the Walker  form \cite{Walker}
\ben
\d s^2=2\d u\left[\d v+H(v,u,x^i)\d u+A_i(u,x^k)\d x^i\right]
+g_{ij}(u,x^k)\d x^i\d x^j,
\label{simmetric}
\een
which is a special case of a Kundt metric\footnote{Note that there exists, 
in general, a coordinate transformation \cite{Ortin} that allows us to set 
$A_i=0$. However, this requires that the transverse metric be $u$-dependent. 
In our case, and especially for the VSI and CSI metrics, it is more useful to 
keep $A_i\neq 0$ so that we can make the transverse metric 
$u$-independent \cite{CSI}.}.

For  four-dimensional  vacuum Kundt spacetimes, the null vector 
$n^\mu$ is, in addition, a thrice repeated  principal null direction of the
Weyl tensor. It follows from a result of Pravda \cite{Pravda1}     
(see also Pravda et al. \cite{Pravda2})
that all invariants formed from the Weyl tensor 
(and since it is Ricci flat, the Riemann tensor) necessarily vanish.

It was pointed out in \cite{Gibbons}
that the vanishing of all invariants  of
 type N vacuum spacetimes with geodesic, non-twisting, non-expanding,
  null congruences implies that all counterterms must vanish,
no  matter what theory of gravity one is considering.
The question was raised in \cite{Gibbons} as to whether this was true for
type III metrics.  This was answered in the affirmative in 
\cite{Pravda1} (see also \cite{Gibbonsloop}).
The question of whether such \lq all loop finite\rq  ~metrics have reduced
holonomy was also raised in \cite{Gibbons}.     In the present  ${\rm Sim}(2)$
Petrov type III 
case we see that, as in the more familar ${\rm E}(2)$ Petrov type N 
case, the answer is yes. However, VSI spaces that do not possess a 
recurrent null vector (the case $\epsilon\neq 0$ in \cite{CFHP}) are 
\lq all loop finite\rq  ~examples having general holonomy.

\section{Boost-weight decomposition} 

In what follows, it is useful to consider the boost-weight decomposition 
of tensors \cite{Milson}. 
Consider an arbitrary covariant tensor $T$ and a null frame
 ${\sf e}^a_{~\mu}=\left\{\ell_{\mu},n_{\mu}, m^{  i}_{\mu}\right\}$; i.e.,
\[ \ell_{\mu}n^{\mu}=1,~~m^{  i}_{~\mu}m^{  j \mu}=
\delta^{  i  j}, ~~ \ell_{\mu}\ell^{\mu}=n_{\mu}n^{\mu}=\ell_{\mu}m^{  i\mu}=
n_{\mu}m^{  i\mu}=0.\]
Consider now a boost in the plane spanned by $\ell^{\mu}$ and $n^{\mu}$:
\ben
{\sf e}^{\tilde{a}}_{~\mu}=\left\{\tilde{\ell}_{\mu},\tilde{n}_{\mu}, 
\tilde{m}^{  i}_{~\mu}\right\}= \left\{e^{\lambda}\ell_{\mu},
e^{-\lambda}n_{\mu}, m^{  i}_{~\mu}\right\}.
\label{boost}\een
We can consider the vector-space decomposition of the tensor $T$ in terms 
of the boost weight with respect to the transformation (\ref{boost})  \cite{Milson}:
\ben
T=\sum_b(T)_b\,,
\een
where $(T)_b$ denotes the projection of the tensor $T$ onto the vector 
space of boost-weight $b$. The components of the tensor $(T)_b$ with 
respect to the null frame will transform according to:
\ben
(T)_{b~ a_1a_2...}=e^{-b\lambda}(T)_{b~\tilde{a}_1\tilde{a}_2...}.
\een
Furthermore, we note that
\ben
\left(T\otimes S\right)_b=\sum_{b'+b''=b}(T)_{b'}\otimes(S)_{b''}.
\een

First, note that the metric has boost-weight decomposition $g=(g)_0$, 
so that raising and lowering indices does not change the boost weight. 
Second, note that any invariant (complete contraction) must be boost 
invariant, so that
\[ {\rm contr}[T]={\rm contr}[(T)_0],\]
where contr means complete contraction 
(i.e., only boost weight 0 components will contribute).

Consider the Riemann tensor $R$ (analogously for the Ricci tensor or the 
Weyl tensor $C$). Recall that the Riemann tensor is algebraically special (otherwise it is of type G)
of a certain type if there exists a frame in which the following hold (in terms of the boost weight 
decomposition) \cite{WeylClass}:
\begin{itemize}
\item{} Type I: $R=(R)_{1}+(R)_0+(R)_{-1}+(R)_{-2}$, 
\item{} Type II: $R=(R)_0+(R)_{-1}+(R)_{-2}$, 
\item{} Type D: $R=(R)_0$, 
\item{} Type III: $R=(R)_{-1}+(R)_{-2}$, 
\item{} Type N: $R=(R)_{-2}$,  
\item{} Type O: $R=0$. 
\end{itemize}

\section{Sim$(n-2)$ VSI and CSI metrics}

We shall consider the $n$-dimensional metric (\ref{simmetric}),
which is a special Kundt metric. For the {\rm Sim}$(n-2)$ metrics there 
are no further restrictions on the metric functions. However, if we make 
the requirement that all the scalar polynomial curvature invariants constructed from the Riemann tensor and its covariant derivatives be constants, then 
the metric function $H(v,u,x^k)$ only contains terms 
polynomial in $v$ to second order \cite{CSI,CFHP}; hence, 
\[ H(v,u,x^k)=v^2\sigma +vH^{(1)}(u,x^k) +H^{(0)}(u,x^k),\]
where   $\sigma$ is a constant. 

Let us choose the null frame: 
\bea
\hat{e}^+&=&\d v+H\d u+A_i\d x^i, \nonumber \\
\hat{e}^-&=&\d u,\nonumber \\
\hat{e}^{i}&=& e^{i}, 
\label{nullframe}\eea
where $e^{i}$ is a vielbein basis for the transverse metric, 
$g_{ij}(u,x^k)\d x^i\d x^j=\delta_{ij}e^i e^j$.  
In general, with respect to the above null frame, the Riemann tensor of  
this metric has the boost weight decomposition
\[ R=(R)_0+(R)_{-1}+(R)_{-2}, \] 
which means it is of Riemann type II 
(and hence, of Ricci type II and of Weyl type II).

\subsection{Vanishing curvature invariants (VSI):}
 
In this case all curvature invariants to all orders vanish, so we require 
that boost-weight zero components vanish \cite{Higher}. This, in turn, 
implies that we can set \cite{CFHP}: 
\[ \sigma=0, \quad g_{ij}\d x^i\d x^j=\delta_{ij}\d x^i \d x^j.\] 
(This is the $W^{(1)}_i=0$ case of the general higher-dimensional VSI metrics.) The Riemann 
tensor is of type III, N or O and has, along with all 
higher-derivative curvature tensors, {\sl only} negative boost weight 
components. 

Considering the vacuum case, 
it must be of Weyl type III. These metrics would therefore be the 
Weyl type III, $\epsilon=0$, vacuum solutions in \cite{CFHP} (see table 1 therein).

\subsection{Constant curvature invariants (CSI):} 

In this case all curvature invariants to all orders are constants. 
Unfortunately, we do not have a crisp theorem determinining all of 
these spacetimes (except in 3 dimensions \cite{3D}), but we have three 
useful \emph{conjectures} \cite{CSI}: 

\begin{enumerate}
\item{} A Lorentzian spacetime with all constant scalar invariants (CSI) 
is either locally homogeneous \emph{or} a subclass of the Kundt spacetimes. 
\item{} For a Kundt spacetime with all constant scalar invariants (CSI), 
there exists a null frame such that for curvature tensors of all orders, 
all the positive boost weight components are zero, while all the boost 
weight 0 components are constants. 
\item{} For a spacetime, $(\mathcal{M},g)$, with all constant scalar 
invariants, there exists a related locally homogeneous spacetime, 
$(\widetilde{\mathcal{M}}_{\text{Hom}},\widetilde{g})$, having identical 
curvature invariants to those of  $(\mathcal{M},g)$. 
\end{enumerate}
We recall that  the class of CSI$_F$ spacetimes are 
those spacetimes for which there exists a null frame such that for 
curvature tensors to all orders, all the positive boost weight components 
are zero, while all the boost weight 0 components are constants. 
The second conjecture therefore suggests that all Kundt CSI spacetimes
are actually CSI$_F$ spacetimes (and consequently, this is usually referred 
to as the CSI$_F$ conjecture). 

Note also that for a Kundt CSI spacetime, this would imply it is of Riemann 
type II (or simpler). It it easy to see that this is sufficient for a 
Kundt spacetime to be CSI.  However, the problematic part of the proof 
is to establish the necessity.  (In three dimensions, all of these 
conjectures have been proven \cite{3D}). 
In what follows, we shall assume these conjectures are true.

An important theorem regarding the CSI$_F$ spacetimes states \cite{CSI}:
\emph{For a Kundt CSI$_F$ there exists a ($u$-dependent) 
diffeomorphism $\hat{x}^i=f^i(u,x^k)$ such that the transverse metric 
can be made $u$-independent. Furthermore, the transverse metric is 
locally homogeneous.} 
Hence, there is no loss of generality in assuming that 
$g_{ij}(x^k)\d x^i\d x^j$ is a locally homogeneous space, $M_{\rm Hom}$. 

Regarding the invariants, there is generally no further simplification; 
they can all be non-zero.
 However, there is a related \emph{locally homogeneous spacetime}, 
$(\widetilde{\cal{M}}_{\rm {Hom}},{\widetilde{g}})$  which has  
invariants that are identical to those of the Kundt CSI metric. 
For the Sim$(n-2)$ metrics, this is given by setting the 
functions $A_i$, $H^{(1)}$ and $H^{(0)}$ to zero: 
\ben
\widetilde{\d s}^2=2\d u\left[\d v+\sigma v^2\d u\right]+
g_{ij}(x^k)\d x^i\d x^j. 
\label{hommetric}\een
For the Kundt CSI metrics with holonomy Sim$(n-2)$, $\sigma$ must be constant. 
Interestingly, this is the space $M_2\times M_{\rm {Hom}}$, 
where $M_2$ is 2-dimensional de Sitter, Minkowski, or anti-de Sitter space, 
depending on the sign of $\sigma$.  

This is an important observation because to study invariants of Kundt CSI 
spacetimes of the form (\ref{simmetric}), it is sufficient 
(provided the conjectures are true) to study the invariants of the associated locally 
homogeneous space (\ref{hommetric}). In particular,
\emph{the locally homogeneous part, defined by (\ref{hommetric}), is the only part 
that contributes to the boost weight 0 components of the curvatures of the 
metric (\ref{simmetric}).} In this way, the invariants will only depend on 
the locally homogeneous metric (\ref{hommetric}). Because of the Kundt 
form, this will also be valid for the higher-derivative curvature tensors. 

\paragraph{Special case: $M_{\rm{Hom}}$ is a symmetric Riemannian space.}
Let us consider the case where the transverse space $g_{ij}(x^k)\d x^i\d x^j$ 
is a symmetric space. This means that the product space (\ref{hommetric}) 
is also symmetric and hence, 
\[ \widetilde{\nabla}_{\lambda}\widetilde{R}_{\mu\nu\alpha\beta}=0.\]
Consequently, for the metric (\ref{simmetric}), \emph{all higher-order 
curvature invariants will vanish.} The only possible non-zero invariants 
are those constructed from the Riemann tensor itself (i.e., the Ricci, Weyl and 
mixed tensor invariants). 

Also note that assuming $g_{ij}$ is a (Riemannian) symmetric space 
implies that $g_{ij}$ is a locally homogeneous space.  

\paragraph{Other examples:}
There are many examples of CSI metrics with Sim$(n-2)$ holonomy (although many are not written down 
explicitly). Assuming the metric is Einstein, we find that the transverse 
metric must also be Einstein. To simplify matters, we can set 
$A_i=0=H^{(1)}$ and still have Sim$(n-2)$ holonomy, provided that 
$H^{(0)}$ is sufficiently general and the transverse metric $g_{ij}(x^k)$ has general holonomy. 
These constitute the case $\beta_i=0$ in \cite{CFH} described in section 3.4. 

Assuming $g_{ij}$ is the metric of a negatively curved space, we can take the transverse 
space to be an Einstein solvmanifold \cite{Heber,solvm1,solvm2}. There are many examples of these, 
and most of them have general holonomy.

\section{Conserved symmetric 2-tensors}

For any 2-tensor we have, in general, the boost weight decomposition: 
\[ T=\sum_{b=-2}^2(T)_b.\]
If this tensor is constructed from contractions of curvature tensors 
from the previous metrics, then there are no positive boost weight 
components. Note that the divergence has the decomposition 
\[{ \rm div}T=({\rm div}T)_1+({\rm div}T)_0+({\rm div}T)_{-1}, \] 
all of which components we require to be zero independently (i.e., they must vanish to each boost weight order separately). 

For the Riemann tensor we also have the Bianchi identity, which upon 
contraction, gives the useful identity:
\[ R^{\alpha}_{~\mu\nu\lambda;\alpha}=R_{\mu\lambda;\nu}-R_{\mu\nu;\lambda}.\]
Furthermore, for a tensor $T_{\alpha_1...\alpha_k}$, we have that
\beq 
\left[\nabla_{\mu},\nabla_{\nu}\right]
T_{\alpha_1...\alpha_k}=T_{\lambda...\alpha_k}
R^\lambda_{~\alpha_1\mu\nu}+\cdots+T_{\alpha_1...\lambda}
R^\lambda_{~\alpha_k\mu\nu}. 
\label{covcom}\eeq
The latter identity implies one may permute the order of covariant 
derivatives of curvature tensors, at the cost of possibly getting products 
of lower-order curvature tensors. 

The various Ricci types and Weyl types restrict the possible non-zero 
tensors constructed from the metric and the curvature tensors. This is perhaps 
best illustrated with examples. 

\subsection{Four dimensions}

\paragraph{Generalised Ghanam-Thompson:} This is a generalisation of a solution found in \cite{Ghanam} and is defined as  a 
4-dimensional Sim$(2)$ Einstein metric, $R_{\mu\nu}=\lambda g_{\mu\nu}$, 
where  $A_i=0$. The Einstein equations imply that $H^{(0)}(u,x^k)$ is 
harmonic in the 2-dimensional space \cite{GibbonsPope}. Regarding the 
Weyl tensor, it is of Weyl (or Petrov) type II: 
\[ C=(C)_0+(C)_{-2}.\] 
 However, the transverse space is the hyperbolic plane \cite{GibbonsPope}, 
and hence symmetric, and so $(\nabla R)_0=0$. As can be checked, 
\[ \nabla^{(k)} R=(\nabla^{(k)} R)_{-2}+(\nabla^{(k)} R)_{-3}+... ~.\]
Any boost weight components of order lower than $-2$ cannot contribute 
to a symmetric 2-tensor, so the only one that can contribute from the 
higher derivatives is $(\nabla^{(k)} R)_{-2}$. (In fact, 
$\nabla^{(k)}R=\nabla^{(k)}C$ since this metric is Einstein.)
It follows that the non-zero invariants are the Ricci scalar, 
zeroth order Weyl invariants, and arbitrary functions of these, $f(I_i)$. 
These are all constants. 
 
The only non-zero symmetric 2-tensors are of the form 
(non-zero boost weight components in brackets):
\begin{itemize}
\item{} $f(I_i)g_{\mu\nu}$ (boost-weight $0$). 
\item{} Products of Weyl tensors with appropriate contractions 
(boost-weight $0$, $-2$)
\item{} $f(I_i) R_{\alpha\beta\gamma\delta;\lambda...}$ with 
appropriate contractions (boost-weight $-2$). Using the contracted 
Bianchi identity, we can show that these are zero too, up to possible 
products of Weyl tensors, which have already been considered. 
\end{itemize}

  At this stage, we have only used boost-weight arguments and the 
Bianchi identites to get possible forms for the conserved 2-tensors. 
However, there are further simplifications. Consider a 2-tensor, 
$T_{\mu\nu}$, made out of an arbitrary product of Riemann tensors. 
This tensor can therefore only consist of  boost-weight $0$ components 
and a boost-weight $-2$ component.  Consider only zeroth order tensors 
first.  The boost-weight $-2$ component is the component $T_{--}$, which can consist of terms: 
\beq 
S^{\mu}_{~\nu}C^{\nu}_{~-\mu-}, \quad S^{\mu}_{~\nu\alpha-}C^{\nu\alpha}_{~~\mu-}, \quad S_{\alpha\beta\mu\nu--}C^{\alpha\beta\nu\mu}, 
\label{SC}\eeq
where $S$ is some tensor consisting of a product of the boost-weight $0$ components of the Riemann tensor $(R)_0$. Consider the first term (with respect to the null-frame (\ref{nullframe})):
\[ T_{--}=S^{\mu}_{~\nu}C^{\nu}_{~-\mu-}=S^{ {i}}_{~ {j}}C^{ {j}}_{~- {i}-}.\]
The tensor $S^{\mu}_{~\nu}$ consists 
entirely of contractions of curvature tensors of the homogeneous 
space $AdS_2\times \mathbb{H}^2$. Hence, by symmetry arguments 
(in the orthonormal frame): 
\[ (S^{\mu}_{~\nu})=\mathrm{diag}(c_1,c_1,c_2,c_2). \] 
Therefore, since $C^{  i}_{~-{  i}-}=0$, we get 
$T_{--}=c_2C^{  i}_{~-{  i}-}=0$. For the other terms in (\ref{SC}), we note that since $S=(S)_0$, any index `$-$' must be accompanied by an index `$+$' (downstairs). Therefore, using symmetry arguments we can show that these vanish too.  Hence, the boost-weight $-2$ component
vanishes. Consider, therefore, the boost-weight $0$ components of $T_{\mu\nu}$: 
these must also consist of a product of $(R)_0$, and hence, must be of the 
same form as $S^{\mu}_{~\nu}$. By a Wick rotation, we can consider the 
Euclidean-signature metric $\mathbb{H}^2\times \mathbb{H}^2$, where the 
curvatures of each $\mathbb{H}^2$ are identical. It is therefore clear 
that\footnote{This can most easily be seen if we note that the isotropy 
group is $U(1)^2\times \mathbb{Z}_2$, where $\mathbb{Z}_2$ interchanges 
the tangent spaces of the $\mathbb{H}^2$. This isotropy group acts 
irreducibly on the tangent space.} $c_1=c_2$, and 
hence, $S_{\mu\nu}=c_1g_{\mu\nu}$. 

     Regarding higher-order invariants, we note that we must consider 
contractions of $(R^{\otimes n})_0\otimes \left(\nabla^{(k)}C\right)_{-2}$. 
We also note that we can write 
\beq (R_{\mu \nu \sigma \tau})_0=c(g^{(1)}_{\mu\sigma}
g^{(1)}_{\nu \tau}-g^{(1)}_{\nu \sigma} 
g^{(1)}_{\mu \tau}+g^{(2)}_{\mu\sigma}
g^{(2)}_{\nu \tau}-g^{(2)}_{\nu \sigma} 
g^{(2)}_{\mu \tau})
\label{eq:R0}\eeq 
where $g^{(1)}$ and $g^{(2)}$ are 2 dimensional projection operators such 
that $g=g^{(1)}\oplus g^{(2)}$. It is therefore also clear that 
contractions of $(R^{\otimes n})_0\otimes \left(\nabla^{(k)}C\right)_{-2}$ 
will vanish.  Combining all of this, we get 
$T_{\mu\nu}=c_1g_{\mu\nu}$, where $c_1$ is some constant. 

We have consequently shown that \emph{the 4-dimensional generalised Ghanam-Thompson solution is 
Weakly Universal. }

\paragraph{Goldberg-Kerr:} This is a 4-dimensional vacuum (Ricci type O) 
and Weyl (or Petrov) type III metric \cite{GoldbergKerr1,GoldbergKerr2}: 
\[ C=(C)_{-1}+(C)_{-2}.\]
Here, all the scalar invariants vanish; i.e., it is a VSI. In general, the 
higher order curvature tensors are: 
\[ \nabla^{(k)} R=(\nabla^{(k)} R)_{-1}+(\nabla^{(k)} R)_{-2}+...\]
(again, it is vacuum so only the Weyl tensor 
will contribute).
Without any further restrictions, we have the following possible 2-tensors:  
\begin{itemize}
\item{} $g_{\mu\nu}$ (boost-weight 0).  
\item{} $R_{\mu\nu} (=0) $, and appropriate contractions of 
$R_{\mu\nu\alpha\beta;\lambda...}$  (boost-weight $-1$, $-2$). 
Using the Bianchi identity these can, without loss of generality, also be assumed to be zero up to 
quadratic terms of (higher-order) curvature tensors (these are considered 
below). 
\item{} Quadratic terms of curvature tensors (boost-weight $-2$). 
\end{itemize}


Let us study these quadratic terms more carefully.  The only non-vanishing 
component of $T_{\mu\nu}$ is the boost-weight $-2$ component $T_{--}$. 
The space is Ricci flat, so only the Weyl tensor will contribute. Consider 
therefore the quadratic term $C_{-\alpha\mu\nu}
\widetilde{C}_{-}^{~\alpha\mu\nu}$, where $C$ and $\widetilde{C}$ can 
be either the Weyl tensor or its dual. Both of these tensors have the same 
symmetries and, in particular, the tracelessness implies 
$C_{- {i}+-}=C^{ {j}}_{~ {i} {j}-}$. In 4 dimensions (and in 4 dimensions only) 
we have as many independent components of $C_{- {i}+-}$ as of 
$C_{ {k} {i} {j}-}$, and by direct calculation we find:
\[ C_{-\alpha\mu\nu}\widetilde{C}_{-}^{~\alpha\mu\nu}=
 -2C_{- {i}+-}\widetilde{C}^{\phantom{-} {i}}_{-~+-}+
  C_{-}^{\phantom{-} {i} {j} {k}}\widetilde{C}_{- {i} {j} {k}}=0.\]
This result can also be derived from the dimensionally dependent identities of Lovelock \cite{Lovelock}, see appendix. 
There could also be a term $C_{-\mu\alpha\nu}
\widetilde{C}_{-}^{~\alpha\mu\nu}$, but this vanishes too for the same reason. Therefore, because of the symmetries of the Weyl tensor, all zeroth order quadratic 
terms will vanish. Each higher-order quadratic term can be written as
\beq
A_{-\alpha\beta\gamma\delta...}B_-^{\phantom{-}\alpha\beta\gamma\delta...},
\eeq
where $A$ and $B$ can be assumed to be linear in the curvature tensors. 
Because of the symmetries of the curvature tensors, the Bianchi identities 
and eq.(\ref{covcom}), we can assume the indices are of a certain order. 
First, this implies that any contraction of $C_{\mu\nu\alpha\beta;\gamma...}$ 
is zero. This further implies that we can, without loss of generality, 
assume that (or let $\alpha\leftrightarrow\beta$ in $B$ )
\beq
A_{-\alpha\beta\gamma\delta...}=C_{-\alpha\beta\gamma;\delta...}, \quad
B_{-\alpha\beta\gamma\delta...}=\widetilde{C}_{-\alpha\beta\gamma;\delta...}.
\eeq 
 Note that $A_{- {i}+-;\delta...}=A^{ {j}}_{~ {i} {j}-;\delta....}$. 
Therefore, consider the ``diagonal'' elements (where $\delta_i$-indices need not be summed): 
\[ A_{-\alpha\beta\gamma{\delta_1}\delta_2...}
B_-^{\phantom{-}\alpha\beta\gamma{\delta_{1}\delta_2}...}=
C_{-\alpha\beta\gamma;{\delta_1}\delta_2...}
\widetilde{C}_-^{\phantom{-}\alpha\beta\gamma;{\delta_1}\delta_2...}=0.\]
Again, this follows from the dimensionally dependent identities of Lovelock \cite{Lovelock} (see appendix).  Thus, all quadratic terms will vanish. 

Consequently, \emph{the 4-dimensional Goldberg-Kerr solution is 
Strongly Universal.}

\paragraph{General 4-dimensional $\mathrm{Sim}(2)$ metrics:}
Indeed, the universality of the Ghanam-Thompson and Goldberg-Kerr solutions extends to all 4-dimensional $\mathrm{Sim}(2)$ metrics:
\begin{thm}\label{thm:sim2}
Consider a 4-dimensional $\mathrm{Sim}(2)$ metric and assume that the 
metric is Einstein, $R_{\mu\nu}=\lambda g_{\mu\nu}$. Then:
\begin{enumerate}
\item{} If $\lambda\neq 0$, the metric is weakly universal. 
\item{} If $\lambda=0$, the metric is strongly universal. 
\end{enumerate}
\end{thm}
This theorem is proven in the appendix and illustrates the remarkable nature of these metrics. 


\subsection{General results in higher dimensions}
We can also summarise some general results for Kundt metrics in higher dimensions in terms of the algebraic properties of the Weyl and Ricci tensor.
\paragraph{Weyl type N, Ricci type O (vacuum):}
 All scalar invariants vanish, all symmetric 2-tensors, except $g_{\mu\nu}$, vanish. 
Therefore, these are  strongly universal. 
\paragraph{Weyl type N, Ricci type N:} 
All scalar invariants vanish. The non-trivial symmetric 2-tensors are $g_{\mu\nu}$, $R_{\mu\nu}$ and 
linear in higher-orders of $R_{\mu\nu}$ (such as, for example, $\Box ^n R_{\mu\nu}$). 
\paragraph{Weyl type N, Einstein space:} 
The only non-vanishing scalar invariant is the Ricci scalar and 
arbitrary functions thereof. The symmetric 2-tensors are thus of the form: $f(I_i)g_{\mu\nu}$. 
These are therefore weakly universal. 

Examples of metrics in this class are the $n$-dimensional `anti-de Sitter waves' with metric 
\[ \d s^2=\frac{1}{b^2z^2}\left[2\d u(\d v+H(u,x^k)\d u)+\d z^2+\delta_{AB}\d x^A\d x^B\right]. \]
These metrics are CSI but do not have holonomy in $\mathrm{Sim}(n-2)$.

\paragraph{Counterexamples:} Here we provide counterexamples in 
higher dimensions which show that not all of the remarkable properties outlined above of the 
4-dimen\-sional Ghanam-Thompson and Golberg-Kerr solutions  extend 
to higher dimensions. 

Consider first the $n$-dimensional analogue of the Ghanam-Thomp\-son solution,
 obtained in \cite{GibbonsPope}.  This Sim$(n-2)$ holonomy
metric can be taken to be 
\beq
\d \hat s^2=2\d u\d v+ \left(H(u,x^k)-2\sigma v^2 \right)\d u^2
+ \d s^2,
\eeq
and it is Einstein, with $\hat R_{\mu\nu}=-2\sigma \hat g_{\mu\nu}$, 
provided that
$\d s^2$ is also Einstein (with $R_{\mu\nu}=-2\sigma g_{\mu\nu}$),
and that $H$ is harmonic over $\d s^2$.  In the vielbein basis $\hat e^+=
  dv+ \ft 12(H-2\sigma v^2) du$, $\hat e^-=du$, $\hat e^i=e^i$, 
the non-zero components of the torsion-free connection and 
the Riemann tensor are given by
\bea
\hat \omega_{+-}&=&2\sigma v \hat e^-\,,\qquad
 \hat\omega_{-i}= \ft12 (\nabla_i H)\, \hat e^-\,,\qquad
\hat\omega_{ij}=\omega_{ij}\,,\\
\hat R_{+-+-}&=& 2\sigma\,,\qquad \hat R_{-i-j}= -\ft12 \nabla_i
\nabla_j H \,,\qquad \hat R_{ijk\ell}=
     R_{ijk\ell}\,. 
\eea
Furthermore, the Weyl tensor decomposes as $C=(C)_0+(C)_{-2}$ in the same 
manner as the 4-dimensional Ghanam-Thompson solution.  Let us consider the
symmetric tensor $\hat S_{ab}\equiv \hat C_{acde} \hat C_b{}^{cde}$. 
If we take the base metric $\d s^2$ to be maximally symmetric, so that
$R_{ijk\ell}=-2\sigma/(n-3) (\delta_{ik}
\delta_{j\ell}-\delta_{i\ell}\delta_{jk})$,
then a simple calculation shows that the non-zero components of $\hat S_{ab}$
are given by
\beq
\hat S_{+-}= \fft{8(n-2)\sigma^2}{n-1}\,,\qquad \hat S_{ij} =
  \fft{16\sigma^2}{(n-1)(n-3)}\,\delta_{ij}\,.
\eeq
It is easily verified that this is divergence-free.
However, it is clearly only in $n=4$ dimensions that it is proportional 
to the metric tensor, and so \emph{this Sim$(n-2)$ Einstein 
spacetime is not weakly universal} in more than four dimensions.

One can also see, by means of counterexamples, that the
generalisations to higher dimensions of the Ricci-flat Goldberg-Kerr 
solutions do not share the strong universality property of the 
four-dimensional case.  The generalised Goldberg-Kerr metric in
$n$ dimensions can be written as
\beq
\d \hat s^2=2\d u \d v + \left(H_0(u,x^k)+ v\, H_1(u,x^k) \right)\d u^2
   + 2 A_i(u,x^k) \d u \d x^i 
+ \d s^2,\label{gkn}
\eeq
where the base metric $\d s^2$ is Ricci flat.  The conditions for 
Ricci-flatness of
$\d\hat s^2$ reduce to \cite{GibbonsPope}
\bea
\nabla^2 H_0 -\ft12 F^{ij} F_{ij} -2A^i \nabla_i H_1 - H_1 \nabla^i A_i
   - 2\nabla^i\dot A_i &=&0\,,\label{gkn1}\\
\nabla^j F_{ij} + \nabla_i H_1 &=&0\,,\label{gkn2}
\eea
where $F_{ij}=\nabla_i A_j -\nabla_j A_i$, and it is assumed that $\d s^2$
is independent of $u$.   One may choose $H_1=-\nabla^i A_i$ as a gauge
choice \cite{GibbonsPope}, in which case Ricci-flatness is achieved by 
first solving the linear equation (\ref{gkn2}) for $A_i$, and
then solving the Poisson equation (\ref{gkn1}) for $H_0$. 

   Let us consider the symmetric tensor 
\beq
\hat T_{ab} = \hat R_{acde} \hat R_b{}^{cde}-
   \ft14 \hat \eta_{ab}\, \hat R_{cdef}
      \hat R^{cdef}\,,
\eeq
which can easily be seen to be conserved, after imposing the condition of
Ricci-flatness.  Consider first the case where $\d s^2$ in (\ref{gkn}) is
the flat Euclidean metric.  We then have
\bea
\hat T_{--} &=& -\ft12 (\nabla_i H_1)(\nabla^i H_1)  + 
 \ft14 (\nabla_i F_{jk})(\nabla^i F^{jk})\,,\nonumber\\
&=& -\ft12(\nabla^i F_{ij})(\nabla_k F^{kj}) + 
      \ft14 (\nabla_i F_{jk})(\nabla^i F^{jk})\,,
\label{eq:T--}\eea
where the second line follows after using (\ref{gkn2}).  For any dimension
$n>4$ this will in general be non-zero, thus showing that there exist
non-vanishing conserved symmetric 2-tensors for these metrics.  Thus, the
generalistion of the Goldberg-Kerr metrics to dimensions higher than
four do not satisfy the strong universality principle.  (The four-dimensional
case is an exception since then we have $F_{ij} = f \epsilon_{ij}$, which
implies that $\hat T_{--}$ vanishes.)
 
We will now consider one specific example where this can be seen explicitly. Consider the $n>4$ dimensional metric with flat transverse metric $\d s^2=\delta_{ij}\d x^i\d x^k$. Furthermore, let 
\beq
A_i\d x^i=xy\d x-azy\d z,
\eeq
where $x,y,z,...$ are Cartesian coordinates on $\d s^2$, and $a$ is a constant. We note that $H_1=-y(1-a)$, so one can verify that eq. (\ref{gkn2}) is indeed satisfied, and eq. (\ref{gkn1}) then reduces to 
\beq
\nabla^2H_0-x^2+y^2(1-a)^2-a^2z^2=0.
\eeq 
This equation has the solution
\beq
H_0=\frac 1{12}\left[x^4-y^4(1-a)^2+a^2z^4\right]+F(u,x^k),
\eeq
where  $F(u,x^k)$ is a solution to $\nabla^2 F=0$. 
Using eq. (\ref{eq:T--}) gives $\hat{T}_{--}=a$ and hence, unless $a=0$, 
this tensor does not vanish. Therefore, \emph{the  solution presented 
above is, for $a\neq 0$,  not strongly universal.}


\paragraph{Semi-universality} 
The (counter)examples of the higher-dimensional Ghanam-Thompson solution naturally leads us to the possibility of another class of metrics with  relatively simple quantum corrections. 

Let us consider the case where the transverse space is a symmetric space 
in which the isotropy group acts irreducibly on the tangent space, 
and define the projection operators:
\[ g^{(1)}=2\hat{e}^+\hat{e}^-, \quad g^{(2)}=
\delta_{ij}\hat{e}^i\hat{e}^j, \] 
so that $g_{\mu\nu}=g^{(1)}_{\mu\nu}+g^{(2)}_{\mu\nu}$. Let us try to 
generalise the Ghanam-Thompson case to higher-dimensions. Therefore, we 
assume the Ricci tensor is of type D and 
\[ R_{\mu\nu}=\lambda_1g^{(1)}_{\mu\nu}+\lambda_2g^{(2)}_{\mu\nu}.\]
We also assume that the Weyl tensor is of the form $C=(C)_0+(C)_{-2}$. Note 
that most of the arguments in the derivation of the symmetric 2-tensors 
of the Ghanam-Thompson solution go through, and we find the following 
general form: 
\beq 
T_{\mu\nu}=f_1(\lambda_1,\lambda_2)g^{(1)}_{\mu\nu}+
f_2(\lambda_1,\lambda_2)g^{(2)}_{\mu\nu}.
\eeq
We note than in  4 dimensions, and for $\lambda_1=\lambda_2$, we can argue 
that $f_1=f_2$. In higher dimensions the equations of motion, including 
quantum corrections, reduce to solving: 
\beq
\lambda_1+f_1(\lambda_1,\lambda_2)=0, \quad 
\lambda_2+f_2(\lambda_1,\lambda_2)=0
\eeq
If such solutions exist, we shall call these metrics 
semi-universal. 

\section{Discussion}

In this paper we have investigated solutions of the classical
Einstein or supergravity equations that remain valid solutions in
the quantum theory. We have been particularly interested in
solutions of the classical Einstein equations which are {\it
universal} ({\it weakly or strongly}), and consequently have a
restricted curvature structure such that they remain solutions of
{\sl almost any} set of covariant equations involving the metric
and its derivatives. An example of such solutions are the
Brinkmann wave class  of solutions of the Einstein equations which
admit a recurrent null vector field \cite{GibbonsPope}, so that
the metrics have holonomy contained in ${\rm Sim} (n-2)$. In
particular, we have focussed attention on the subsets of  Einstein
metrics which are VSI  or CSI.

In four dimensions we have found two  new classes of metrics, one
of which is strongly universal and the other of which is weakly
universal. In more detail, in the 4-dimensional generalised
Ghanam-Thompson spacetime, which is a Sim$(2)$ Einstein CSI metric
(with $A_i=0$ and a harmonic $H^{(0)}(u,x^k)$ in which the  Weyl
tensor is of type II), the only non-zero invariants are the Ricci
scalar, zeroth order Weyl invariants, and arbitrary functions of
these, which are all constants. By a direct calculation we then
found that the only non-zero conserved symmetric 2-tensors are of
the form $c g_{\mu\nu}$, where $c$ is a constant. Therefore, the
4-dimensional generalised Ghanam-Thompson solution is \emph{weakly
universal}. In the 4-dimensional vacuum and Weyl type III
Goldberg-Kerr VSI metric, all of the scalar invariants vanish. By
a direct calculation we then found that the metric (modulo a
constant rescaling) is the only non-zero conserved symmetric
2-tensor. Consequently, the 4-dimensional Goldberg-Kerr solution
is \emph{strongly universal.}

We also showed that \emph{all} 4-dimensional ${\rm Sim}(2)$ Einstein metrics are universal, and  in the future it will be of interest to consider  4-dimensional Einstein metrics with general holonomy and investigate under what circumstances such metrics are weakly universal ($\lambda\neq
0$) or strongly universal ($\lambda=0$). Since the Goldberg-Kerr and the Ghanam-Thompson solutions are of different algebraic type there is hope that their properties will extend to an even bigger class of metrics.

We then discussed \emph{universality} in higher dimensional
spacetimes. We first summarised some general results in higher
dimensions and discussed some examples. In particular, we noted
that the higher dimensional generalizations of the Ghanam-Thompson
and Golberg-Kerr solutions are not (weakly and strongly,
respectively) universal. It is therefore clear that Theorem \ref{thm:sim2} cannot be generalised to higher-dimensions. It was suggested that the notion of
\emph{semi-universality} may be of relevance in studying  the
quantum corrections of higher-dimensional classical solutions. In
future we hope to further study the higher dimensional case.

\section*{Acknowledgements}
SH would like to thank N Pelavas for helpful discussions regarding the proof of Theorem \ref{thm:sim2}. 
This work was supported by NSERC (AAC and SH). CNP is supported in part 
by DOE grant DE-FG03-95ER40917.

\appendix
\section{Proof of Theorem \ref{thm:sim2}} 
Here we intend to prove: \\
{\it Consider a 4-dimensional $\mathrm{Sim}(2)$ metric and assume that the 
metric is Einstein, $R_{\mu\nu}=\lambda g_{\mu\nu}$. Then:
\begin{enumerate}
\item{} If $\lambda\neq 0$, the metric is Weakly Universal. 
\item{} If $\lambda=0$, the metric is Strongly Universal. 
\end{enumerate}
}
\paragraph{Dimensionally dependent identities:} In four dimensions, and four dimensions only, we have the identity \cite{Lovelock}: 
\beq
\delta^{[\mu}_{~[\nu}{C^{\alpha\beta]}}_{\delta\gamma]}=0.
\eeq
This identity is what Lovelock calls a dimensionally dependent identity and is valid for a Weyl tensor, or any Weyl-like tensor with the same symmetries as the Weyl tensor. By considering covariant derivatives of this identity we obtain therefore:
\[ \delta^{[\mu}_{~[\nu}{C^{\alpha\beta]}}_{\delta\gamma];\epsilon_1\dots\epsilon_k }=0.\] 
Consider a tensor $W_{\alpha\beta\delta\gamma\epsilon_1\dots\epsilon_k}$, where 
\[W_{\alpha\beta\delta\gamma\epsilon_1\dots\epsilon_k}=W_{[\alpha\beta]\delta\gamma\epsilon_1\dots\epsilon_k}=W_{\delta\gamma\alpha\beta\epsilon_1\dots\epsilon_k}, \quad W^{\alpha}_{~\beta\alpha\gamma\epsilon_1\dots\epsilon_k}=0. \]
By contraction, we get the identity: 
\beq
0=g_{\mu\nu}\left(W^{\alpha\beta\gamma\delta\epsilon_1\dots\epsilon_k}C_{\alpha\beta\gamma\delta;\eta_1\cdots\eta_k}\right)+4{W_{\alpha(\nu}}^{\beta\gamma\epsilon_1\dots\epsilon_k}C^{\phantom{\mu}\alpha}_{\mu)\phantom{\alpha}\beta\gamma;\eta_1\dots\eta_k}.
\label{eq:TCid}\eeq
Note that if we are considering the diagonal components (i.e., $\epsilon_1=\eta_1,\dots,\epsilon_k=\eta_k$), then the boost weight of the first term in eq.(\ref{eq:TCid}) is zero. Hence, the second term can only have boost weight 0 terms also. This identity very useful in proving univerality for $\mathrm{Sim}(2)$ metrics. 
\paragraph{Einstein equations:} For a $\mathrm{Sim}(2)$ metric the Einstein equations imply that
\[ H(v,u,x^k)=v^2\lambda +vH^{(1)}(u,x^k) +H^{(0)}(u,x^k),\]
and the transverse metric, $g_{ij}(u,x^k)$ is a 2-dimensional Einstein space. 
In particular, this means that the transverse space is symmetric and 
either $S^2$, $\mathbb{E}^2$, $\mathbb{H}^2$. Moreover, the $\mathrm{Sim}(2)$-metric 
must be CSI$_F$ (VSI if $\lambda=0$) and we can use a coordinate 
transformation to get rid of the $u$-dependence in the transverse 
metric (i.e., $g_{ij}(x^k)$). 
\paragraph{Boost weight decomposition:} 
Regarding the zeroth order curvature tensors, these are of Ricci type D  and Weyl (Petrov) type II (for $\lambda=0$ it is of type III):
\[ C=(C)_0+(C)_{-1}+(C)_{-2}. \] 
Since the transverse metric is symmetric, higher-order curvature tensors have the following boost weight decomposition ($k>0$):
\[ \nabla^{(k)} C=\left(\nabla^{(k)} C\right)_{-1}+ \left(\nabla^{(k)} C\right)_{-2}+...\] 
For the symmetric tensor $T_{\mu\nu}$ we have that
\[ T=(T)_0+(T)_{-1}+(T)_{-2},\]
where each boost-weight component will be considered in turn. 
\paragraph{boost-weight $0$:} 
The boost-weight $0$ components of $T_{\mu\nu}$ are identical to those 
of the symmetric space $\widetilde{\d s}^2=2\d u\left[\d v+\lambda 
v^2\d u\right]+g_{ij}(x^k)\d x^i\d x^j$. Considering the isotropy group, 
we can see that any curvature 2-tensor must be of the form 
$S^{\mu}_{~\nu}=\mathrm{diag}(c_1,c_1,c_2,c_2)$. Since the curvatures 
match, we must have $c_1=c_2$, and hence 
\[ S_{\mu\nu}=c_1g_{\mu\nu}.\] 
\paragraph{boost-weight $-1$:} These components must arise as contractions 
of terms like 
\[ (R^{\otimes n})_{0}\otimes(C)_{-1}, \quad 
(R^{\otimes n})_{0}\otimes\left(\nabla^{(k)}C\right)_{-1}. \] 
First, consider zeroth order terms. In this case it is advantageous to switch to the Weyl canonical frame. This will not change the b.w. 0 terms but might alter the negative boost weight terms. For $\lambda\neq 0$, the spacetime is of Weyl type II, and in the Weyl canonical frame we have $C=(C)_0+(C)_{-2}$. Therefore, in this frame the b.w. $-1$ components are zero. For $\lambda=0$ (which implies Ricci flat) the Weyl canonical frame gives $C=(C)_{-1}$ and therefore zeroth order terms give no contributions to the b.w. $-1$ components of the symmetric tensor $T_{\mu\nu}$.

Next, consider the higher order terms $(R^{\otimes n})_{0}\otimes\left(\nabla^{(k)}C\right)_{-1}$. We easily see that the number of covariant derivatives must be even in order for this term to contribute to the quantum corrections. Now, the boost weight 0 curvature tensor has the form eq. (\ref{eq:R0}) which implies that $(R^{\otimes n})_{0}$ must be a tensor product of $g^{(1)}$ and $g^{(2)}$.
For the $\mathrm{Sim}(2)$ metrics we note that the components $C_{\alpha\beta\gamma\delta;\epsilon_1\dots\epsilon_k}=0$ if there is an $\epsilon_i$-index being `+'  (or else, the holonomy would not be (a subgroup of) $\mathrm{Sim}(2)$). Therefore, if $g^{(1)}=2\hat{e}^+\hat{e}^-$, we have $C_{\alpha\beta\gamma\delta;\epsilon_1\dots\epsilon_i\dots\epsilon_j\dots\epsilon_k}g^{(1)\epsilon_i\epsilon_j}=0$. Thus, since $g=g^{(1)}+g^{(2)}$, 
\[ C_{\alpha\beta\gamma\delta;\epsilon_1\dots\epsilon_i\dots\epsilon_j\dots\epsilon_k}g^{(2)\epsilon_i\epsilon_j}=C_{\alpha\beta\gamma\delta;\epsilon_1\dots\epsilon_i\dots\epsilon_j\dots\epsilon_k}g^{\epsilon_i\epsilon_j}.\]
By permuting the indices $\epsilon_i$ and $\epsilon_j$ we can show, up to lower-order products, that this must vanish too.  
Consequently, any contraction of $g^{(2)}$ with $\left(\nabla^{(k)}C\right)_{-1}$ is also zero. Since  $(R^{\otimes n})_{0}$ is a tensor product of $g^{(1)}$ and $g^{(2)}$,  $(R^{\otimes n})_{0}\otimes\left(\nabla^{(k)}C\right)_{-1}$ cannot contribute to a symmetric 2 tensor. 

Hence, we are lead to the conclusion that all b.w. $-1$ components must vanish.  
 
\paragraph{boost-weight $-2$:}
This component contains contractions of terms like: 
\[ (R^{\otimes n})_{0}\otimes(C)_{-2}, \quad 
(R^{\otimes n})_{0}\otimes\left(\nabla^{(k)}C\right)_{-2}, \quad 
\left(\nabla^{(k_1)}C\right)_{-1} \otimes\left(\nabla^{(k_2)}C\right)_{-1}. \] 
Contributions from the first and second terms must vanish (this follows 
from an identical argument as for the Ghanam-Thompson solution).  The 
vanishing of the third term follows from permuting the indices using eq. (\ref{covcom}), using the Bianchi identity, and using eq. (\ref{eq:TCid}).

\end{document}